\documentclass[10pt,aps,prc,twocolumn,superscriptaddress,floatfix]{revtex4-1}
\usepackage{graphicx}
\usepackage{amsmath}
\usepackage{braket}
\usepackage{url}
\usepackage{xcolor}
\usepackage{multirow}
\usepackage{amssymb} 
\usepackage{booktabs}
\usepackage{rotating}
\usepackage{bm}
\usepackage{bbm}
\usepackage{longtable}
\usepackage{subfigure}

\usepackage[colorlinks,
linkcolor=blue,
anchorcolor=red,
urlcolor=red,
citecolor=red]{hyperref}

\usepackage[utf8]{inputenc}
\usepackage[T1]{fontenc}

\usepackage[left]{lineno}
\begin{document}

\title{Experimental determination of the double $K$-shell vacancy creation probability in the electron-capture decay of $^{55}$Fe with a microcalorimeter}

\affiliation{State Key Laboratory of Heavy Ion Science and Technology, Institute of Modern Physics, Chinese Academy of Sciences, Lanzhou 730000, China} 
\affiliation{Narodowe Centrum Badań Jądrowych, Andrzeja Sołtana 7, 05-400 Otwock-Świerk, Poland} 
\affiliation{University of Chinese Academy of Sciences, Beijing 100049, China} 
\affiliation{China Center for Transformative Science, ShanghaiTech University, Shanghai 201210, China} 
\affiliation{School of Nuclear Science and Technology, Lanzhou University, Lanzhou 730000, China} 

\author{Shuo Zhang} \email[]{shuozhang@impcas.ac.cn}
\affiliation{State Key Laboratory of Heavy Ion Science and Technology, Institute of Modern Physics, Chinese Academy of Sciences, Lanzhou 730000, China}
\affiliation{China Center for Transformative Science, ShanghaiTech University, Shanghai 201210, China}

\author{Karol Kozioł}
\affiliation{Narodowe Centrum Badań Jądrowych, Andrzeja Sołtana 7, 05-400 Otwock-Świerk, Poland}

\author{Caojie Shao}\email[]{c.shao@impcas.ac.cn}
\affiliation{State Key Laboratory of Heavy Ion Science and Technology, Institute of Modern Physics, Chinese Academy of Sciences, Lanzhou 730000, China}
\affiliation{University of Chinese Academy of Sciences, Beijing 100049, China}

\author{Ruitian Zhang}
\affiliation{State Key Laboratory of Heavy Ion Science and Technology, Institute of Modern Physics, Chinese Academy of Sciences, Lanzhou 730000, China}
\affiliation{University of Chinese Academy of Sciences, Beijing 100049, China}

\author{Cheng Qian}
\affiliation{State Key Laboratory of Heavy Ion Science and Technology, Institute of Modern Physics, Chinese Academy of Sciences, Lanzhou 730000, China}
\affiliation{University of Chinese Academy of Sciences, Beijing 100049, China}

\author{Kebao Shu}
\affiliation{School of Nuclear Science and Technology, Lanzhou University, Lanzhou 730000, China}
\affiliation{State Key Laboratory of Heavy Ion Science and Technology, Institute of Modern Physics, Chinese Academy of Sciences, Lanzhou 730000, China}

\author{JingKai Xia}
\affiliation{China Center for Transformative Science, ShanghaiTech University, Shanghai 201210, China}

\author{Guocheng Wei}
\affiliation{State Key Laboratory of Heavy Ion Science and Technology, Institute of Modern Physics, Chinese Academy of Sciences, Lanzhou 730000, China}
\affiliation{University of Chinese Academy of Sciences, Beijing 100049, China}

\author{Yingli Xue}
\affiliation{State Key Laboratory of Heavy Ion Science and Technology, Institute of Modern Physics, Chinese Academy of Sciences, Lanzhou 730000, China}
\affiliation{University of Chinese Academy of Sciences, Beijing 100049, China}

\author{Huan Jia}
\affiliation{State Key Laboratory of Heavy Ion Science and Technology, Institute of Modern Physics, Chinese Academy of Sciences, Lanzhou 730000, China}
\affiliation{University of Chinese Academy of Sciences, Beijing 100049, China}

\author{Neng Pu}
\affiliation{State Key Laboratory of Heavy Ion Science and Technology, Institute of Modern Physics, Chinese Academy of Sciences, Lanzhou 730000, China}
\affiliation{University of Chinese Academy of Sciences, Beijing 100049, China}

\author{Hanjie Cai}
\affiliation{State Key Laboratory of Heavy Ion Science and Technology, Institute of Modern Physics, Chinese Academy of Sciences, Lanzhou 730000, China}
\affiliation{University of Chinese Academy of Sciences, Beijing 100049, China}

\author{Deyang Yu}
\affiliation{State Key Laboratory of Heavy Ion Science and Technology, Institute of Modern Physics, Chinese Academy of Sciences, Lanzhou 730000, China}
\affiliation{University of Chinese Academy of Sciences, Beijing 100049, China}

\author{Hongqiang Zhang}
\affiliation{School of Nuclear Science and Technology, Lanzhou University, Lanzhou 730000, China}

\author{Liangting Sun}
\affiliation{State Key Laboratory of Heavy Ion Science and Technology, Institute of Modern Physics, Chinese Academy of Sciences, Lanzhou 730000, China}
\affiliation{University of Chinese Academy of Sciences, Beijing 100049, China}

\author{Yuan He}
\affiliation{State Key Laboratory of Heavy Ion Science and Technology, Institute of Modern Physics, Chinese Academy of Sciences, Lanzhou 730000, China}
\affiliation{University of Chinese Academy of Sciences, Beijing 100049, China}

\author{Reinhold Schuch}
\affiliation{Physics Department, Stockholm University, Stockholm S-10691, Sweden}

\author{Xinwen Ma}
\affiliation{State Key Laboratory of Heavy Ion Science and Technology, Institute of Modern Physics, Chinese Academy of Sciences, Lanzhou 730000, China}
\affiliation{University of Chinese Academy of Sciences, Beijing 100049, China}

\date{\today}

\begin{abstract}
The probability of double \(K\)-shell vacancy production following electron-capture decay is determined for $^{55}$Fe using a high-resolution microcalorimeter. The \(K\)-hypersatellite lines arising from double \(K\)-shell-vacancy filling are directly resolved within the measured spectra of X-ray emissions which are dominated by Mn \(K\)-X-ray diagram lines. The energies of the Mn \textit{K}-hypersatellite lines, \(K\alpha_2^h\), \(K\alpha_1^h\), and \(K\beta_{1,3}^h\), are measured as $(6143.5\pm0.5)$ eV, $(6162.1\pm1.5)$ eV and $(6811.9\pm2.6)$ eV, respectively. The relative intensities of \textit{K}-hypersatellite to diagram X-rays are measured to  be $(1.30 \pm 0.07) \times 10^{-4}$. From the relative intensities together with established fluorescence yield values, the probability of double-vacancy creation per \(K\)-shell capture ($P_{{KK}}$) is determined as $(1.22 \pm 0.09) \times 10^{-4}$. Comparisons with previous experimental results and theoretical calculations are discussed.

\end{abstract}


\maketitle

\section{INTRODUCTION}

In a radioactive decay via \textit{K}-shell electron capture (EC) of a nucleus, the atom is typically left with a single \textit{K}-shell vacancy. The sudden change of nuclear charge, coupled with the disappearance of the captured electron, can cause the remaining \textit{K}-shell electron to be excited to unoccupied bound levels (shake-up) or ejected into the continuum (shake-off), thereby creating a double \textit{K}-shell vacancy. Although this double-vacancy production process occurs with a very low probability (on the order of $10^{-6}$ to $10^{-2}$), it has attracted considerable scientific interest. This interest stems not only from its high sensitivity to electron correlation effects between the two \textit{K}-shell electrons in the atom's initial state in various areas of physics~\cite{McGuire_1997}, but also from the unique opportunity it provides to investigate angular momentum coupling~\cite{CUE1977,Diamant2003,Chantler2014} and exotic decay modes in this distinctive regime~\cite{Wolfli1975,Salem1982,Mokler1990,FOLKERTS1992}.

Since the first observation of double-vacancy production during \textit{K}-electron capture in the decay of $^{55}$Fe by Charpak \textit{et al.} using gaseous detectors in a coincidence experiment~\cite{Charpak1953}, extensive research has been conducted to measure the probabilities (\textbf{$P_{KK}$}) of this process per \textit{K}-shell electron capture across a broad range of EC nuclides with atomic numbers ($Z$) from 18 to 83. An overview of measured values for different nuclei is provided in the work of Hindi and Kozub~\cite{Hindi1992}. Most of these measurements reasonably agree with the theoretical treatment of Primakoff and Porter (PP), who predicted a $1/Z^2$ behavior of \textbf{$P_{KK}$}~\cite{Primakoff1953}.

However, comparative analysis of $P_{KK}$ values reveals significant deviations among the neighboring nuclides: $^{54}$Mn ($Z=25$), $^{65}$Zn ($Z=30$), and $^{55}$Fe ($Z=26$). The experimental $P_{KK}$ values for $^{54}$Mn reported by Hindi \textit{et al.}~\cite{Hindi2003} and for $^{65}$Zn by Nagy and Schupp~\cite{Nagy1983} are $(2.3^{+0.8}_{-0.5})\times 10^{-4}$ and $(2.2 \pm 0.2) \times 10^{-4}$, respectively. These values are found to substantially exceed the corresponding measurements for $^{55}$Fe, where the experimental $P_{KK}$ values are $(1.2 \pm 0.4) \times 10^{-4}$ reported by Briand \textit{et al.}~\cite{BRIAND1974} and $(1.3 \pm 0.2) \times 10^{-4}$ by Campbell \textit{et al.}~\cite{Campbell1991}. In response, a progressive increase in precision has been demonstrated in recent measurements of the $P_{KK}$ value for $^{55}$Fe. Michel \textit{et al.}~\cite{Michel2014} reported $(1.531 \pm 0.079) \times 10^{-4}$, and Bergmann \textit{et al.}~\cite{Bergmann2016} obtained a more precise value of $(1.388 \pm 0.037)\times 10^{-4}$. These somewhat larger results point to the resolution of the discrepancy regarding the $1/Z^{2}$ scaling of $P_{KK}$ for $^{55}$Fe relative to $^{54}$Mn, but not relative to $^{65}$Zn.

With the exception of the study by Campbell \textit{et al.}~\cite{Campbell1991}, all measurements of $P_{KK}$ in electron-capture decays have been conducted using two or more detectors in coincidence mode. This is because the double \textbf{$K$}-vacancy created by EC decay is rapidly filled by electrons from higher shells. This process results in the subsequent emission of secondary particles, such as two photons, two Auger electrons, or one photon accompanied by one Auger electron. Consequently, the double-vacancy state can be identified by the simultaneous detection of these particle pairs using high-efficiency detectors, such as high-purity germanium (HPGe) detectors positioned around the radioactive source. Generally, coincidence measurements require sophisticated electronic systems to precisely determine the temporal correlation between events and correct for random coincidence backgrounds, which significantly increases system complexity and introduces additional uncertainties.

\begin{figure}[htbp]
\centering
\includegraphics[width=\linewidth]{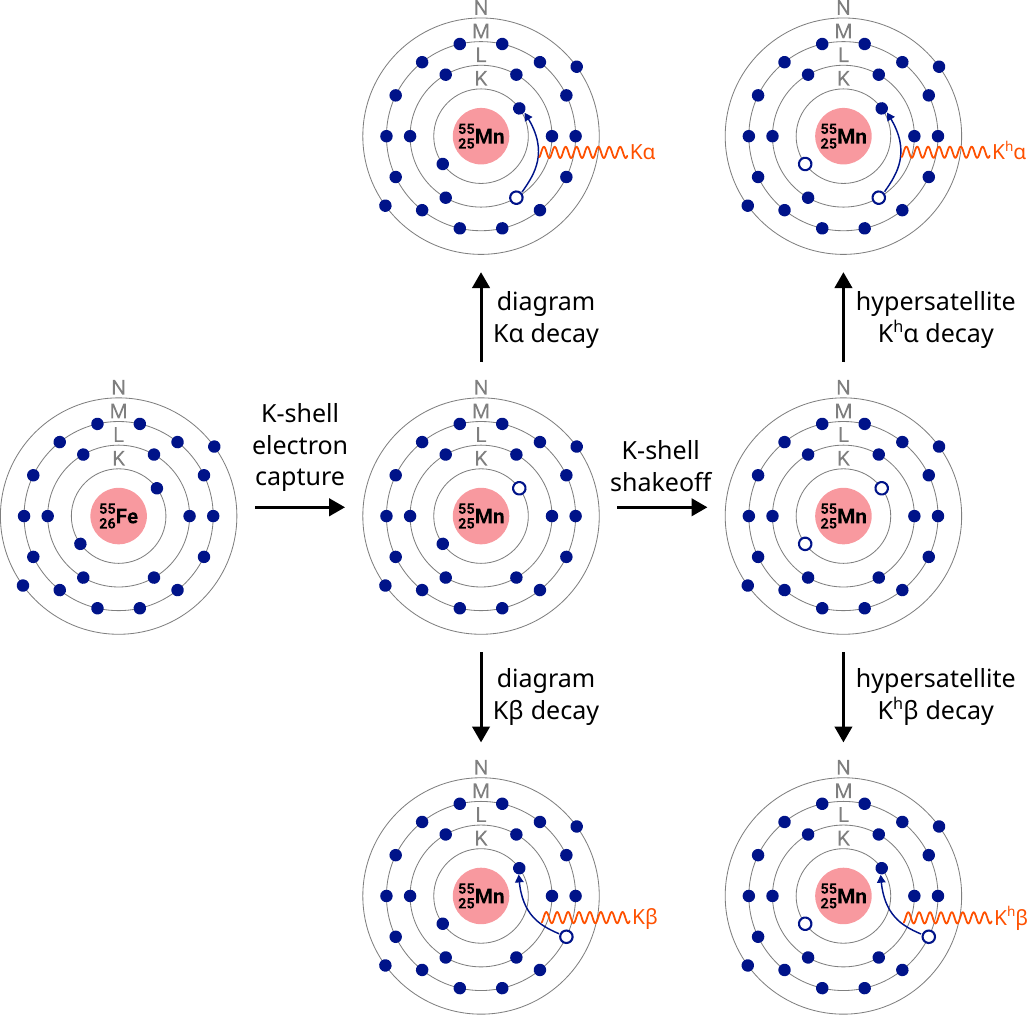}

\caption{Schematic of \textsuperscript{55}Fe EC decay forming \textsuperscript{55}Mn with a \textit{K}-shell vacancy, followed by \textit{K}-shell shake-off and radiative de-excitation pathways.}
\label{fig1}
\end{figure}

Due to the reduced screening of the nuclear charge caused by an additional \textit{K}-shell vacancy, the X-ray lines emitted from atoms with double \textit{K}-shell vacancies—known as \textit{K}-hypersatellite lines (see Fig.~\ref{fig1})—are shifted to higher energies compared to diagram lines from atoms with a single \textit{K}-shell vacancy. Briand \textit{et al.} were the first to clearly demonstrate this effect, showing that the energies of gallium X-ray hypersatellite lines in the EC decay of $^{71}$Ge shift to higher energies relative to \textit{K}-diagram lines~\cite{Briand1971}. Subsequently, \textit{K}-hypersatellite lines from atoms across a wide range of atomic numbers, produced either by the EC process or by collisions of target atoms with ions and photons~\cite{Richard1972,Olsen1974,Diamant2003,RZADKIEWICZ1999,Rzadkiewicz2000,Hoszowska2009,Fennane2009,Diamant2009,Diamant200902,Hoszowska2010,Tamasaku2013,Shao2017,Hillenbrand2022,Song2025}, have been measured and well distinguished using a single high-energy-resolution detector, and the corresponding $P_{KK}$ values have been obtained and summarized. It is noteworthy, however, that the only measurement of $P_{KK}$ on an EC-decay nuclide using a single-detector setup was conducted by Campbell \textit{et al.}~\cite{Campbell1991}, who employed a Si(Li) detector to record the X-ray spectrum of $^{55}$Mn from the EC decay of $^{55}$Fe. In this experiment, the energy resolution of the detector used was comparable to the energy difference between the \textbf{$K\alpha$} hypersatellite and the diagram line. Along with the intrinsically low $P_{KK}$ value it resulted in the $K\alpha$ hypersatellite remaining invisible in the energy spectra, necessitating the use of a sophisticated spectral fitting method.

Through analysis of previous results, Campbell has pointed out that measurements from high-resolution X-ray spectrometers yield significantly lower $P_{KK}$ values than earlier low-resolution methods and that such measurement are presumably less susceptible to interference from radioactive contaminants. To the best of our knowledge, the measurement of $P_{KK}$ for an EC decay nuclide through the X-ray spectrum with a high-energy-resolution ($E/\Delta E \sim 1000$) detector has not been carried out before our experiment. This is due to the fact that energy-dispersive high-resolution detectors, such as crystal spectrometers have low efficiencies which means high-activity radioactive sources or long measurement time are required.

The tremendous progress in development of high-resolution, efficient energy-deposition detectors i.e. microcalorimeters~\cite{McCammon2005,Irwin2005,Fleischmann2005} has opened new avenues for studying double \textit{K}-shell vacancy production in EC decaying nuclei. With such a microcalorimeter, we were capable of detecting low-intensity hypersatellites, enabling simultaneous measurement of diagram and well resolved hypersatellite lines, to address the discrepancy observed in $P_{KK}$ of very similar atomic numbers. The main goal is to explore the double \textit{K}-vacancy creation probability in manganese and the subsequent vacancy filling process following $^{55}$Fe EC decay. The paper is organized as follows. Section II provides a brief description of the experiment. Section III presents the analysis of the X-ray spectra. Results and discussions are presented in Sec.~IV, and finally a summary of the present work and a brief outlook are given in Sec. V.

\section{EXPERIMENT}
The radioactive source was prepared by HTA Co., Ltd. via electrodeposition of $^{55}$Fe onto a copper substrate (radius 12.5 mm, thickness 2 mm). The certified activity of the source was $3.7 \times 10^8$ Bq on January 4, 2019. The emitted X-rays from the radioactive source passed through a layer of 25 $\mu$m low-density polyethylene film and were detected by a microcalorimeter at a distance of 150 mm from the source surface. The radiation source and the detector were shielded with 6 mm thick lead to minimize background from environmental photons. 

\begin{figure*}[htbp]
\centering
\includegraphics[width=\linewidth]{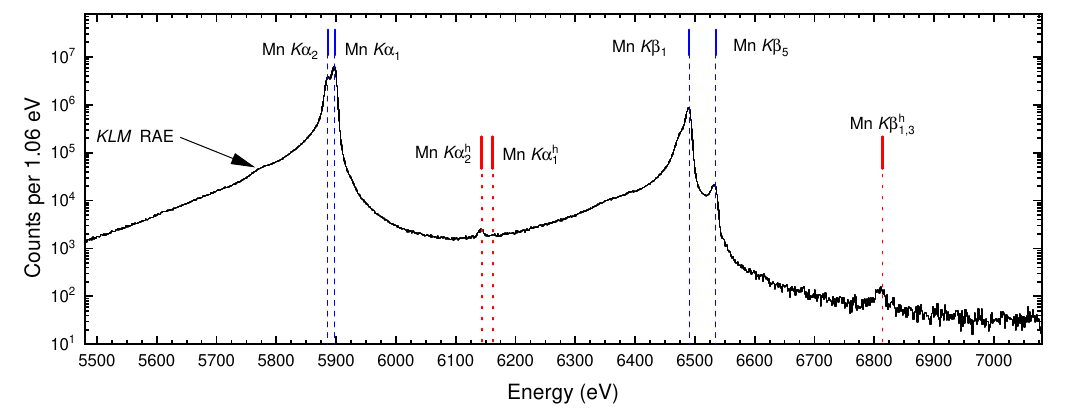}
\caption{Manganese \( K \) X-ray spectrum from \( ^{55}\)Fe decay recorded using a microcalorimeter. The positions of the \( K\alpha_{1,2}, K\beta_{1} \) and \( K\beta_5 \) diagram lines, as well as the weak \( K\alpha_2 \) and \( K\beta_{1,3} \) hypersatellite lines (\( K\alpha_2^h \) and \( K\beta_{1,3}^h \)), are indicated by vertical markers.  \(KLM\) RAE indicates the position of the radiative Auger effect.}
\label{fig2}
\end{figure*}

The present X-ray detector is a superconducting transition-edge-sensor (TES) microcalorimeter. The details of this microcalorimeter were described elsewhere~\cite{Zhang2022, Wu2023}. Briefly, the detectors use a $4 \times 4$ pixel microcalorimeter sensor chip, and the area of a single pixel is approximately $0.5 \text{ mm} \times 0.5 \text{ mm}$, thus resulting in a relative solid angle of $\Delta\Omega/\Omega$ being $1.1 \times 10^{-5}$ for each pixel. The thermal sensor is a molybdenum-copper bilayer film with a transition temperature of approximately 70 mK with molybdenum leads. The absorber layer is a 2.4 $\mu$m-thick bismuth layer with an intrinsic detection efficiency of 70\% at 5.89 keV and 61\% at 6.49 keV. A 3-$\mu$m aluminum-plated Mylar film and thin IR/vis/UV blocking filters were placed in front of the pixels for thermal insulation and to block low-energy photons. This TES detector can detect photons with energies ranging from 0.6 to 20 keV. To ensure energy resolution, the count rate of each pixel was kept below 20 counts per second during the whole measurement period. A total of 80 million counts were acquired in the current experiment. The energy spectrum obtained from each pixel was independently analyzed and processed.

\section{X-ray SPECTRA ANALYSIS}

The spectrum of Mn X-rays from the EC decay of $^{55}$Fe, recorded by the TES microcalorimeter, is presented in Fig.~\ref{fig2}, with the positions of the Mn $K\alpha_1$, $K\alpha_2$, $K\beta_{1}$, and $K\beta_{5}$ diagram lines indicated by vertical lines. Besides these intense diagram lines, two distinct spectral features are evident: one peak situated between the $K\alpha$ and $K\beta$ groups and another on the high-energy side of the $K\beta_{5}$ line. The energies of these additional peaks serve to rule out the possibility of their origin being diagram lines from impurity atoms. We also indicate a small hump below Mn $K\alpha_{1,2}$ by $KLM$ RAE, which is due to the Radiative Auger Effect (RAE), a process where the $K$-excitation energy is shared between a photon and an ejected electron.

In the present study, the detector's energy scale was self-calibrated with the $^{55}$Fe source based on four reference peaks: Mn $K\alpha_1$ at 5898.10 eV, Mn $K\beta_{1,3}$ at 6490.26 eV, Cu $K\alpha_2$ at 8028.38 eV, and Cu $K\alpha_1$ at 8048.11 eV~\cite{Deslattes2003} (the latter induced by background radiation in the TES copper heatsink), using a second-order polynomial to calibrate the energy-to-channel relationship. Each peak is fitted to line profiles using the Bortels function~\cite{BORTELS1987} to model the detector's energy response. This function is defined as the convolution of a Gaussian with a normalized sum of a low-energy one-sided exponential tail and a delta function. The exponential tail originates from the incomplete thermalization of X-ray energy in the evaporated bismuth absorber of each TES~\cite{Daikang,Neil2020}. Each spectral peak is characterized by five parameters: the centroid and area of the Gaussian component, along with the detector response characteristics, specifically comprising the Gaussian width $\sigma$, the tail fraction $f_{\text{tail}}$, and the tail scale length $\lambda$~\cite{JParc2022}. 
To enhance the fit quality and adjust to the spectral complexity, the fitting procedure is done in four segments, with slightly
varying parameters in each segment, as explained below.

\begin{figure*}[htbp]
\centering
\includegraphics[width=\linewidth]{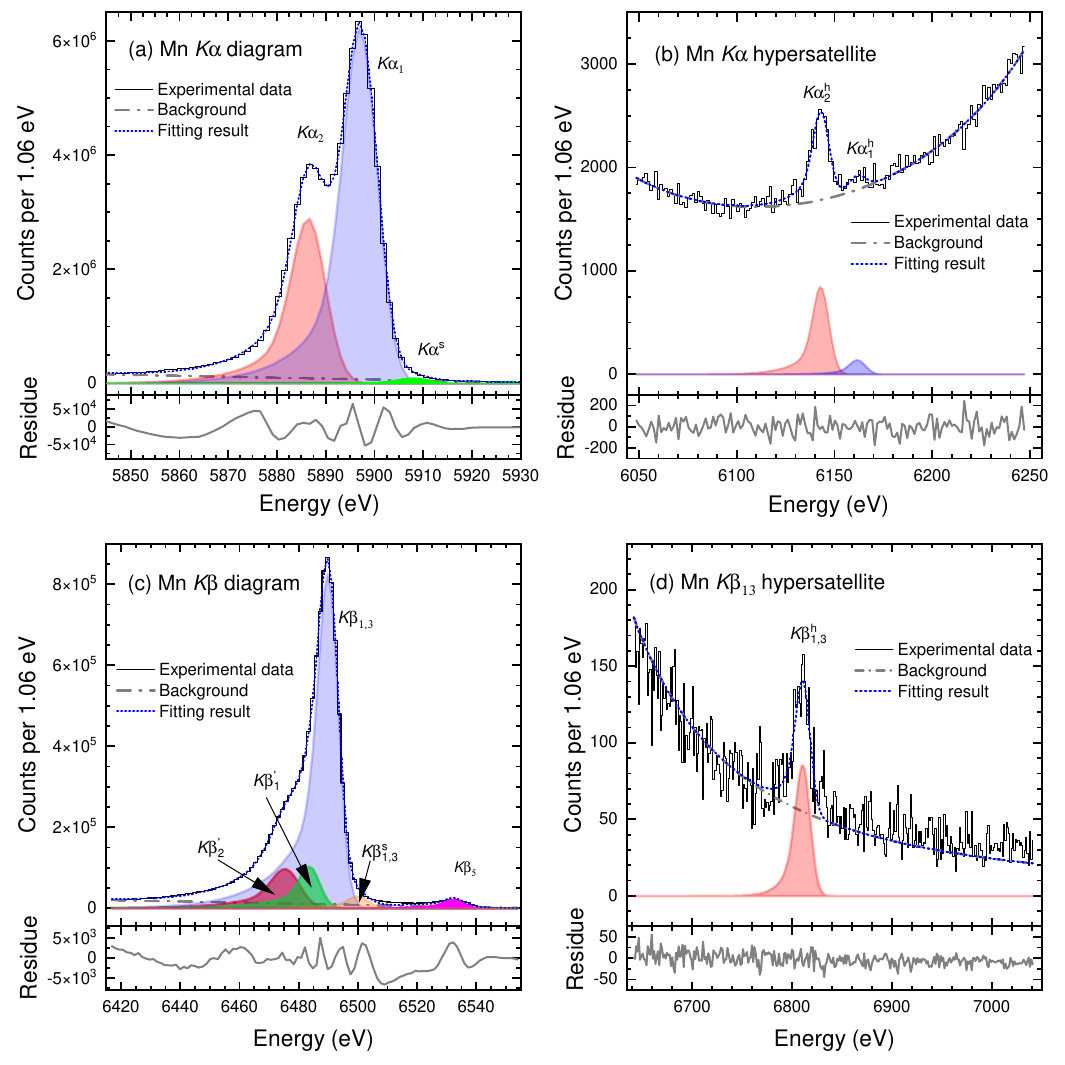}
\caption{Details of the spectral fits for \( ^{55}\)Fe: (a) \( K{\alpha_{1,2}} \) diagram, (b) \( K{\alpha_{1,2}} \) hypersatellite, (c) \( K{\beta} \) diagram, and (d) \( K{\beta_{1,3}} \) hypersatellite lines. All fits were performed with the Bortels function following prior background subtraction. The residuals of each fit are shown in the corresponding subpanels.}
\label{fig3}
\end{figure*}

For the \textbf{$K\alpha$} and \textbf{$K\beta$} diagram line segments, the fitting range for each was defined relative to the centroids of their constituent peaks, spanning from 12.8$\sigma$ below the lowest-energy peak to 6.3$\sigma$ above the highest-energy peak (where $\sigma$ is the Gaussian width parameter). This specific asymmetric interval was determined through numerical integration of the line shape to ensure it encompasses the vast majority ($>99.999\%$) of each spectral component's total area, thereby preventing truncation errors in intensity extraction. This selection also provides a stable, featureless baseline\textcolor{red}{,} crucial for reliable background determination. For the \textbf{$K\alpha_{1,2}$} diagram line complex, substantial residuals on the high-energy side of the \textbf{$K\alpha_1$} line were observed during the two-peak fitting. These discrepancies are attributed to the neglect of $K\alpha$ satellite lines, which originate from \textit{K}-\textit{L} radiative decay processes in atoms with additional \textit{L}- or \textit{M}-shell vacancies. Hence, this part of the spectrum \textbf{is} fitted with three Bortels distributions following linear background subtraction, and the parameters $\sigma$, $f_{\text{tail}}$, and $\lambda$ are each assigned common values across these closely spaced peaks.

The \textbf{$K\beta$} X-ray spectrum comprises the $K\beta_{1,3}$ and $K\beta_5$ diagram peaks, accompanied by a satellite structure. The satellite on the low-energy side of the $K\beta_{1,3}$ peak is collectively termed \textbf{$K\beta'$}, while the one on the high-energy side is referred to as $K\beta_{1,3}^{\mathrm{s}}$. The origin of this $K\beta'$ satellite has been particularly studied, with proposed mechanisms including exchange interaction~\cite{Salem1976} and plasmon oscillations~\cite{Srivastava1979}. The $K\beta_{1,3}^{\mathrm{s}}$ line arises from a transition that occurs in the presence of a spectator vacancy in the \textit{L}- or \textit{M}-shell. In our analysis, we adopted the approach of Ito \textit{et al.} for analyzing high-resolution Mn spectra induced by an X-ray tube, where the low-energy satellite is modeled as two distinct peaks: \textbf{$K\beta'_1$} and \textbf{$K\beta'_2$}~\cite{Ito2018}. Since their measured energy splitting between $K\beta_1$ and $K\beta_3$ is only 1.4 eV---significantly less than the detector's resolution---these lines are treated as a single peak. Accordingly, the $K\beta$ spectrum was fitted with five Bortels functions after background removal, with shared values for the $\sigma$, $f_{\text{tail}}$, and $\lambda$ parameters, except for the $\sigma$ of the lowest-energy peak \textbf{$K\beta'_2$}. This exception allows for potential broadening of RAE~\cite{Limandri2010}. Thus, its $\sigma$ is fitted independently. Details of the spectral fits for $^{55}$Fe are shown in Fig.~\ref{fig3}\textbf{~}(a) for $K\alpha$ diagram lines and (c) for $K\beta$ diagram lines.

For the \textbf{$K\alpha$} hypersatellite lines, the background underlying the peaks is evaluated by curve-fitting to adjacent data points and then removed. The peaks are subsequently fitted with two Bortels distributions using the same parameter set and constraints as applied to the $K\alpha_1$ and $K\alpha_2$ diagram lines. For the \textbf{$K\beta$} hypersatellite lines, following the same background evaluation and removal procedure, the peaks are fitted with a single Bortels distribution. The fitting results for the \textbf{$K\alpha$} and \textbf{$K\beta$} hypersatellite lines are presented in Fig.~\ref{fig3}\textbf{~}(b) and (d), respectively.

The peak energies obtained through above fitting procedure and their corresponding uncertainties, are compiled and presented in Table~\ref{table1}. The obtained energy resolutions, full width at half maximum (FWHM), were 6.84 eV at 5.898 keV and 7.09 eV at 6.490 keV. For comparison, previously reported experimental ~\cite{BRIAND1974,CUE1977,Rahkonen1977,Diamant2009,Jonnard2002} and theoretical ~\cite{ABERG1982,Chen1982,Diamant2009,Wang201293} values, are also summarized in Table~\ref{table1}. Measured intensity ratios of the \textit{K}-hypersatellite lines to their corresponding \textit{K}-diagram lines, as well as the intensity ratios of the $K\beta$ hypersatellite to the $K\alpha$ hypersatellite line $I(K\beta^h)/I(K\alpha^h)$ and of the $K\alpha_1$ to the $K\alpha_2$ hypersatellite lines $I(K\alpha_{1}^{h})/I(K\alpha_{2}^{h})$, are compiled and presented in Table~\ref{table2}. The relative intensities $I(K\beta_{1,3}^{h})/I(K\beta)$ and $I(K\beta_{1,3}^{h})/I(K\beta_{1,3})$ correspond to cases where the $K\beta$ diagram lines are considered inclusive of and exclusive of the $K\beta'_2$, $K\beta^s$, and $K\beta_5$ spectral components, respectively. The ratio $I(K^{\mathrm{h}})/I(K)$ is derived using the exclusive definition of the $K\beta$ diagram lines.

\begin{table}[htbp]
\caption{
Experimental and theoretical energies of the \textit{K}-hypersatellite lines \(K{\alpha_{2}^h}\), \( K{\alpha_{1}^h}\), and \( K{\beta_{1,3}^h}\), along with the satellite lines \(K{\alpha_{}^s}\), \(K{\beta_{}^s}\), $K\beta'_1$ , and  $K\beta'_2$. The uncertainty is given by the fitting standard error. See text for details.
}
\centering
\begin{tabular}{lccc}
\hline\hline
\multirow{2}{*}{Lines} 
& \multicolumn{1}{c}{Present} & \multicolumn{1}{c}{Previous} & \multicolumn{1}{c}{Theoretical} \\
& \multicolumn{1}{c}{(eV)} & \multicolumn{1}{c}{(eV)} & \multicolumn{1}{c}{(eV)} \\
\hline
$K\alpha_{2}^{\mathrm{h}}$ 
& $6143.5 \pm 0.5$ 
& 
\begin{tabular}{c}
$6143.4\pm0.1$\textsuperscript{a}\\
$6142.0\pm3.0$\textsuperscript{b}\\
$6143.2\pm3.0$\textsuperscript{c}\\
\end{tabular}
&

\begin{tabular}{c}
6140.9\textsuperscript{a}\\
6143.0\textsuperscript{d}\\
\\
\end{tabular}
\\
\hline

$K\alpha_{1}^{\mathrm{h}}$ 
& $6162.1 \pm 1.5$ 
& 
\begin{tabular}{c}
$6160.9\pm0.1$\textsuperscript{a}\\
$6162.2\pm3.0$\textsuperscript{c}
\end{tabular}
&

\begin{tabular}{c}
6158.4\textsuperscript{a}\\
6161.3\textsuperscript{d}\\
\end{tabular}
\\
\hline

$K\beta_{1,3}^{\mathrm{h}}$ 
& $6811.9 \pm 2.6$ 
& $6835\pm35$\textsuperscript{e}
& 6813.7\textsuperscript{f}\\
\hline

$K\alpha^{\mathrm{s}}$ 
& $5908.3 \pm 0.7$ 
&
& 5910.0\textsuperscript{\textbf{h}} \\
\hline

$K\beta'_1$ 
& 
\begin{tabular}{c}
$6484.1 \pm 0.5$ \\
\end{tabular}
& $6485.45 \pm 0.23$\textsuperscript{g} 
&  \\
\hline

$K\beta'_2$
& 
\begin{tabular}{c}
$6476.5 \pm 0.6$ \\
\end{tabular}
& $6478.30\pm 0.19$\textsuperscript{g} 
&   \\
\hline

$K\beta^{\mathrm{s}}$ 
& $6500.7 \pm 0.3$   
&-&-  \\
\hline
\hline

\multicolumn{4}{p{8.5cm}}{$^{a}$Diamant \textit{et al}., Ref.~\cite{Diamant2009}; \hspace{0.22em}  $^{b}$Keski-Rahkonen \textit{et al}., Ref.~\cite{Rahkonen1977};} \\
\multicolumn{4}{p{8.0cm}}{$^{c}$Cue \textit{et al}.,
Ref.~\cite{CUE1977};\hspace{2.9em} $^{d}$
\r{A}berg \textit{et al}., Ref.~\cite{ABERG1982};   } \\
\multicolumn{4}{p{8.5cm}}{ $^{e}$Briand \textit{et al}., Ref.~\cite{BRIAND1974};\hspace{1.0em} $^{f}$Chen \textit{et al}., 
Ref.~\cite{Chen1982};} \\
\multicolumn{4}{p{8.5cm}}{$^{g}$Jonnard \textit{et al}., Ref.~\cite{Jonnard2002}; \hspace{0.0em} $^{h}$Wang \textit{et al}., 
~\cite{Wang201293}.} \\

\end{tabular}
\label{table1}
\end{table}

\begin{table*}[htbp]
\begin{minipage}{.99\textwidth}
\caption{Experimentally determined intensity ratios involving \textit{K}‑hypersatellite lines, listed before and after corrections for self‑absorption and detection efficiency. The uncertainty is given by the fitting standard error. See text for details.}
\centering
\renewcommand{\arraystretch}{1.25}
\begin{tabular}{llcccc}
\hline
Relative Intensity
&Before  correction
& 
\begin{tabular}{c}
Correction  factor \\
\textit{f} \\
\end{tabular}

& After  correction 
& Previous result 
&Present  theory \\
\hline
$I(K\alpha_{1,2}^{\mathrm{h}})/I(K\alpha_{1,2})$
& $(1.29\pm0.08)\times10^{-4}$ 
& 1.017
&$(1.27\pm0.08)\times10^{-4}$ \\
$I(K\beta_{1,3}^{\mathrm{h}})/I(K\beta)$
& $(1.31\pm0.11)\times10^{-4}$ 
& 1.018
& $(1.29\pm0.10)\times10^{-4}$ 
& $(1.30\pm0.20)\times10^{-4}$\textsuperscript{,a}\\
$I(K\beta_{1,3}^{\mathrm{h}})$/$I(K\beta_{1,3})$ 

&$(1.60\pm0.14)\times10^{-4}$ 
& 1.018
&$(1.57\pm0.14)\times10^{-4}$ \\
$I(K^{\mathrm{h}})/I(K)$
& 
&  
& $(1.30\pm0.07)\times10^{-4}$ \\

$I(K\beta_{1,3}^{\mathrm{h}})/I(K\alpha_{1,2}^{\mathrm{h}})$ 
& $0.144 \pm 0.015$
& 1.041
& $0.138 \pm 0.014$ 
&
&0.134 \textsuperscript{b}\\
$I(K\alpha_{1}^{\mathrm{h}})/I(K\alpha_{2}^{\mathrm{h}})$
&  
&  
& $0.164 \pm 0.038$ 

&
\begin{tabular}{c}
$0.19\pm0.05$\textsuperscript{c}\\
$0.13\pm0.01$\textsuperscript{d}\\
\end{tabular}
&
 0.155 \textsuperscript{b} \\
\hline
\multicolumn{6}{p{16.0cm}}{$^{a}$Campbell \textit{et al}., Ref.~\cite{Campbell1991}.       
$^{b}$Koziol and Rzadkiewicz, Ref.~\cite{KOZIOL2020}. 
$^{c}$Cue \textit{et al}., Ref.~\cite{CUE1977}.
$^{d}$Diamant \textit{et al}., Ref.~\cite{Diamant2009}.}

\end{tabular}
\label{table2}
\end{minipage}
\end{table*}

When determining $P_{KK}$ from the relative intensity of the $\textit{K}\alpha$ hypersatellite to the diagram line, it is necessary to account for the self-absorption factors of the X-rays within the radiative source, as well as for their relative detection efficiencies. The experimental X-ray intensity ratios $I(\textit{K}\beta)/I(\textit{K}\alpha)$, $I(\textit{K}\alpha^{h})/I(\textit{K}\alpha)$ and $I(\textit{K}\beta^{h})/I(\textit{K}\alpha)$ were derived from the measured spectra using the following expressions: \begin{equation} \frac{I(\textit{K}\beta)}{I(\textit{K}\alpha)} = \frac{N(\textit{K}\beta)}{N(\textit{K}\alpha)} \frac{\rho(\textit{K}\alpha)}{\rho(\textit{K}\beta)} \frac{\epsilon(\textit{K}\alpha)}{\epsilon(\textit{K}\beta)} \end{equation} \begin{equation} \frac{I(\textit{K}\alpha^{h})}{I(\textit{K}\alpha)} = \frac{N(\textit{K}\alpha^{h})}{N(\textit{K}\alpha)} \frac{\rho(\textit{K}\alpha)}{\rho(\textit{K}\alpha^{h})} \frac{\epsilon(\textit{K}\alpha)}{\epsilon(\textit{K}\alpha^{h})} \end{equation} \begin{equation} \frac{I(\textit{K}\beta^{h})}{I(\textit{K}\alpha)} = \frac{N(\textit{K}\beta^{h})}{N(\textit{K}\alpha)} \frac{\rho(\textit{K}\alpha)}{\rho(\textit{K}\beta^{h})} \frac{\epsilon(\textit{K}\alpha)}{\epsilon(\textit{K}\beta^{h})} \end{equation} where $N(\textit{K}\alpha)$, $N(\textit{K}\beta)$, $N(\textit{K}\alpha^{h})$, and $N(\textit{K}\beta^{h})$ are the net counts for the $\textit{K}\alpha$, $\textit{K}\beta$, $\textit{K}\alpha^{h}$, and $\textit{K}\beta^{h}$ peaks, respectively; $\rho(\textit{K}\alpha)$, $\rho(\textit{K}\beta)$, $\rho(\textit{K}\alpha^{h})$, and $\rho(\textit{K}\beta^{h})$ are the source self-absorption correction factors for these peaks; and $\epsilon(\textit{K}\alpha)$, $\epsilon(\textit{K}\beta)$, $\epsilon(\textit{K}\alpha^{h})$, and $\epsilon(\textit{K}\beta^{h})$ are the corresponding detection efficiencies. In a given thin-source experiment, the self-absorption factor and the detection efficiency for $\textit{K}$ hypersatellite lines, each normalized to the corresponding values for the $\textit{K}\alpha$ line, depend solely on energy. This dependence can be considered approximately linear in the narrow 5.89--6.90 keV range, thus permitting their values to be obtained via linear extrapolation from a single $\textit{K}\beta$-to-$\textit{K}\alpha$ intensity ratio measurement.

Specifically, the relative self-absorption correction factor $\rho(\textit{K}\beta)/\rho(\textit{K}\alpha)$ was determined by comparing the measured intensity ratio $I(\textit{K}\beta)/I(\textit{K}\alpha)$ from the present $^{55}$Fe source with the reference value of 0.1359 provided by Sch\"onfeld \textit{et al.}~\cite{SCHONFELD1996}. This reference value agrees well with our independent measurement on a thin, nearly self-absorption-free source, which yields 0.136. The experimental ratio was measured to be 0.146 using a Silicon Drift Detector (SDD) at near-zero source-to-detector distance, ensuring nearly identical detection efficiencies for both lines. This yields a correction factor of 1.074. For the hypersatellites detected at 6.14 keV ($\textit{K}\alpha^{h}$) and 6.81 keV ($\textit{K}\beta^{h}$), the relative self-absorption factors were derived as $\rho(\textit{K}\alpha^{h})/\rho(\textit{K}\alpha) = 1.03$ and $\rho(\textit{K}\beta^{h})/\rho(\textit{K}\alpha) = 1.11$, respectively. It is important to note that both the literature values and our SDD measurements rely on Gaussian modeling, where integrated areas inherently include satellite lines and partial RAE contributions. Since these spectral features are common to both datasets, the comparison-based determination of the self-absorption correction factor remains valid and unaffected by these structures.

The relative detection efficiencies for the $\textit{K}\beta$ to $\textit{K}\alpha$ diagram lines, $\varepsilon(\textit{K}\beta)/\varepsilon(\textit{K}\alpha)$, are governed by their attenuation over the 150-mm path through air and various foils, as well as the intrinsic efficiency of the calorimeter for these lines. This efficiency ratio was determined by comparing the intensity ratio measured by the TES with the reference value of 0.146 obtained from SDD measurements in a close-contact geometry. Given that the TES-measured $K\beta$-to-$K\alpha$ intensity ratio is 0.140, the resulting efficiency ratio is $\varepsilon(\textit{K}\beta)/\varepsilon(\textit{K}\alpha) = 0.956$. Consequently, the detection efficiencies of the TES for the $\textit{K}\beta^{h}$ and $\textit{K}\alpha^{h}$ lines relative to that of the $\textit{K}\alpha$ line are also derived via linear extrapolation as $\varepsilon(\textit{K}\alpha^{h})/\varepsilon(\textit{K}\alpha) = 0.981$ and $\varepsilon(\textit{K}\beta^{h})/\varepsilon(\textit{K}\alpha) = 0.930$, respectively.

Through the analytical method described above, we obtained the total relative correction factor $f_i = (\rho_i \varepsilon_i)/(\rho_0 \varepsilon_0)$, with subscripts $i = 0, 1, 2$, and 3 that correspond to the  $\textit{K}\alpha_{1,2}$, $\textit{K}\alpha_{1,2}^{h}$, $\textit{K}\beta_{1,3}$, and $\textit{K}\beta_{1,3}^{h}$ lines. This factor incorporates corrections for both the self-absorption and detection efficiency of the various spectral lines relative to the $\textit{K}\alpha$ line. With the $\textit{K}\alpha$ line ($i=0$) as the reference ($f_0 \equiv 1.000$), the values for the other lines were found to be $f_1 = 1.017$, $f_2 = 1.039$, and $f_3 = 1.058$. Following the relative intensity results from the spectral fitting, their corrected values (accounting for self-absorption and detection efficiency) were determined. These corrected values, together with the corresponding correction factors, are summarized in Table~\ref{table2}.

\section{RESULTS AND DISCUSSION} 

As shown in Table~\ref{table1}, the measured energies of the manganese $K\alpha_{2}^{h}$ and $K\alpha_{1}^{h}$ lines are in agreement with previous high-resolution values, both for photon-induced~\cite{Diamant2009,Rahkonen1977} and EC-decay experiments~\cite{CUE1977}. They also agree with the calculations of \AA berg~\cite{ABERG1982} to within $\le 0.7$~eV, while being larger than that of Diamant \textit{et al}.~\cite{Diamant2009} by $\sim 3$~eV. The calculated value by Diamant et al. was also confirmed to be lower then their own measured value. The energies of the manganese $K\beta_{1,3}^{h}$ lines are obtained here for the first time with an uncertainty of ±2.6 eV, and are consistent with theoretical predictions by Chen \textit{et al}~\cite{Chen1982}, differing by only 1.8 eV.  The measured energy shifts of the $K\alpha^{\mathrm{s}}$  and $K\beta^{\mathrm{s}}$ lines relative to their respective diagram lines are both on the order of 10 eV. According to theoretical calculations by Wang \textit{et al.}~\cite{Wang201293}, the average shift induced by an \textit{L}‑shell spectator vacancy is about 28 eV for $K\alpha^{\mathrm{s}}$ and 65 eV for $K\beta^{\mathrm{s}}$, accounting for the various possible configurations. This suggests that both the  $K\alpha^{\mathrm{s}}$  and $K\beta^{\mathrm{s}}$ lines are more likely to originate from transitions involving a spectator vacancy in the \textit{M}‑shell, rather than one in the \textit{L}‑shell.

The measured $K\beta_{1,3}^{h}$-to-$K\alpha_{1,2}^{h}$ intensity ratio of $0.138 \pm 0.014$ is found to agree well with our theoretical predictions of $0.134$~\cite{KOZIOL2020} (see Table~\ref{table2}). The $K\alpha_{1}^{h}$-to-$K\alpha_{2}^{h}$ ratio, which is the most sensitive indicator for tracking angular momentum coupling variations across the periodic table, is measured to be $0.164 \pm 0.038$. This value agrees with our theoretical prediction of $0.155$, and is slightly higher than the previous photon-induced result of $0.13 \pm 0.01$~\cite{Diamant2009, Rahkonen1977} and somewhat lower than the internal-conversion result of $0.19 \pm 0.05$~\cite{CUE1977}. Despite these slight differences, it remains fully consistent with all previous values within the experimental uncertainties.

The production probability of a double \textit{K}-shell vacancy per \textit{K}-shell electron capture decay, $P_{{KK}}$, can be determined using the following formula:
\begin{equation}
P_{KK}
=
\frac{I(K^{h})/\omega_{K}}{I(K)/\omega_{KK}}
=
\frac{I(K^{h})}{I(K)} \frac{\omega_{KK}}{\omega_{K}}
\tag{4}
\end{equation}
where $\omega_{K}$ and $\omega_{KK}$ are the $K$ X-ray fluorescence yields for the single and double $K$-vacancy states of the Mn atom, respectively. 
For consistency with the earlier analysis by Bergmann \textit{et al.}~\cite{Bergmann2016}, we adopted the same \(K\)-shell fluorescence yield of \(\omega_K = 0.321\) for manganese, originally reported by Schönfeld \textit{et al.}~\cite{SCHONFELD1996}, a choice also supported by the agreement between our independently measured $K\beta/K\alpha$ intensity ratio and theirs as mentioned in Section III. Similarly, to maintain consistency with the reference data set used by Bergmann, we employed the same double \textit{K}-shell fluorescence yield of $\omega_{KK} = 0.344$, a value originally reported by Chen \textit{et al}~\cite{Chen1991} and independently verified by the calculations~\cite{KOZIOL2020} which yielded a very similar value of 0.340, thereby confirming its reliability for the current study. Combining the relative X-ray emission intensities from Table~\ref{table2} with the fluorescence yield ratio $\omega_{KK}/\omega_{K} = 1.07$, and incorporating a detection efficiency and fluorescence uncertainty of $<2\%$, we determined the double \textit{K}-vacancy production probability for $^{55}\mathrm{Fe}$ to be $P_{KK} = (1.22 \pm 0.09)\times 10^{-4}$.

\begin{figure}[htbp]
\centering
\includegraphics[width=\linewidth]{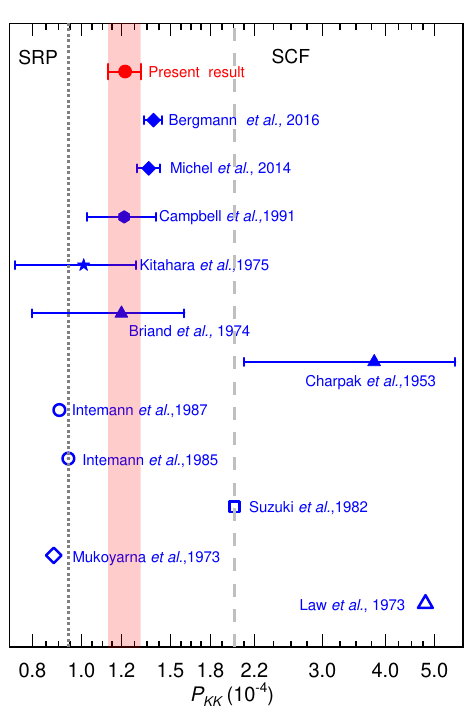}
\caption{Comparison of measured $P_{KK}$ value of $^{55}$Fe ($Z$=25) with theoretical and others' experimental values. Experiment: {\color{red}$\bullet$} present result, {\color{blue}$\blacklozenge$} coincident measurement with pixel detectors~\cite{Michel2014,Bergmann2016}, {\color{blue}$\blacksquare$} by single energy detector~\cite{Campbell1991}, {\color{blue}$\bigstar$} by measuring Shake-off electrons~\cite{Kitahara1975}, {\color{blue}$\blacktriangle$} by coincident measurement~\cite{Charpak1953}. Theory: {\color{blue}$\bigcirc$} Correlation-split wave function~\cite{Intemann1987}, {\color{blue}$\odot$} Semi-relativistic Coulomb propagator method (SRP)~\cite{Intemann1985}, {\color{blue}$\square$} SCF potential plus final-state correlation using the local density approximation (SCF)~\cite{Suzuki1982}, {\color{blue}$\diamond$} Initial state correlation by screening constant~\cite{Mukoyama1973}, {\color{blue}$\triangle$} Correlation effects neglected~\cite{Law1973}.}
\label{fig4}
\end{figure}

For comparison, our experimental result is presented alongside existing theoretical predictions and prior experimental data for $^{55}\text{Fe}$ in Fig.~\ref{fig4}. It should be noted that the values quoted from Michel \textit{et al.}~\cite{Michel2014} and Campbell \textit{et al.}~\cite{Campbell1991} have been adjusted according to the analysis by Bergmann \textit{et al.}~\cite{Bergmann2016}. Specifically, the Michel \textit{et al.} result was corrected to $P_{KK} = (1.358 \pm 0.070)\times 10^{-4}$ by adopting higher fluorescence yields for both $\omega_{KK}$ and $\omega_{KL}$, while the Campbell \textit{et al.} result was revised to $P_{KK} = (1.2 \pm 0.2) \times 10^{-4}$ by correcting for $\omega_{KK}$ under the assumption of unchanged $K\beta$ fractional intensities between single- and double-vacancy. Here $\omega_{KL}$ is the fluorescence yields for filling a single $K$-shell vacancy in the presence of a spectator vacancy in the $L$-shell. As illustrated in Fig.~\ref{fig4}, aside from the earliest data point by Charpak, the $P_{KK}$ value obtained in the present work is consistent with these corrected results and earlier studies; however, it remains significantly lower than the recent value $(1.388 \pm 0.037) \times 10^{-4}$ reported by Bergmann \textit{et al}~\cite{Bergmann2016}. 

To understand this discrepancy, we first performed a comparative analysis focusing on the two experiments that utilized a single-sided energy detection setup. In the study by Campbell \textit{et al}.~\cite{Campbell1991}, the limited energy resolution of the detector prevented extraction of the $K\alpha$ hypersatellite signal, leading to determination of $P_{KK}$ solely from the $K\beta^{\mathrm{h}}$-to-$K\beta$ line intensity ratio, with a reported value of $(1.3 \pm 0.2)\times 10^{-4}$. Furthermore, the $K\beta$ diagram line obtained in that work was in fact an integrated spectral feature encompassing the $K\beta_{1,3}$, $K\beta_{1}'$, $K\beta_{2}'$, $K\beta^{\mathrm{s}}$ and even $K\beta_{5}$ components without individual resolution. Accordingly, our analysis adopts this identical treatment of the $K\beta$ diagram line complex. The resulting ratio of $(1.29 \pm 0.10) \times 10^{-4}$ shows excellent agreement with the value reported by Campbell \textit{et al}.

The inclusion of $K\beta'_1$ and $K\beta'_2$ lines as components of the $K\beta$ diagram line depends on whether their physical origin affects the fluorescence yield. If these lines arise from exchange interactions or plasmon oscillations---processes corresponding to energy level shifts in the \textit{M} shell---they would not significantly alter the fluorescence yield~\cite{BE19981367}. Conversely, if they originate from the radiative Auger effect, they should be excluded from the fluorescence yield definition. Measurements of the Mn $K\beta$ line under photon excitation, as reported by Y.~Ito \textit{et al}., indicate that the $K\beta'_2$ line displays a pronounced line width characteristic of RAE. In our measurements, its relative intensity compared to the $K\beta_{1,3}$ line is 18\%. Therefore, in our analysis, we exclude $K\beta'_2$ along with the $K\beta_{1,3}L^{-1}$ satellite and $K\beta_{5}$ lines from the $K\beta_{1,3}$ diagram line spectral feature, retaining only $K\beta_1'$ as part of the $K\beta_{1,3}$ line. This approach yields a $K\beta_{1,3}^h$-to-$K\beta_{1,3}$ intensity ratio of $(1.57 \pm 0.14)\times 10^{-4}$, which is significantly higher than results obtained when all components are included, and seems to be consistent with the findings reported by Bergmann \textit{et al}.

Compared to the $K\beta$ line, the $K\alpha$ line exhibits higher intensity and a spectral profile that is less affected by outer-shell electrons, resulting in a much simpler structure with far fewer spectral components. The \textit{KLM} RAE lines, located on the low-energy side of the $K\alpha$ diagram lines, exhibit low intensity ($\sim$0.5\% relative to $K\alpha$) in both theory and experiment~\cite{Scofield1974,Campbell1986}. In this work, their peak is observed at approximately 5780 eV (Figure~\ref{fig2}), distinct from the $K\alpha$ diagram lines. The $K\alpha L^{-1}$ satellite line on the high-energy side is resolved via the fitting procedure described in Section II and has an intensity of $\sim$1.06\% relative to the $K\alpha$ diagram lines. After subtracting the contributions of the two satellite structures, the obtained $K\alpha^h$-to-$K\alpha$ line intensity ratio is $(1.27 \pm 0.08)\times 10^{-4}$. This leads to the  $K^h$-to-$K$  line intensity ratios with the $K\beta'_2$ line either included in or excluded from the $K\beta$ diagram lines being $(1.28 \pm 0.07)\times 10^{-4}$ or $(1.30 \pm 0.07)\times 10^{-4}$, respectively. The very small difference between these values thus supports the greater reliability of the $P_{KK}$ value derived from the $K^h$-to-$K$ line intensity ratio compared to that obtained solely from the $K\beta^h$-to-$K\beta$  line intensity ratio.

Assuming $P_{KK}$ follows the $1/Z^2$ dependence predicted by the PP theory~\cite{Primakoff1953}, and using the measurement of $^{54}\text{Mn}$ by Hindi \textit{et al}. together with the fluorescence-yield-corrected extrapolation to $^{54}\text{Cr}$ reported by Bergmann \textit{et al}., one obtains $P_{KK}(^{55}\text{Fe}) = (1.87^{+0.71}_{-0.41})\times 10^{-4}$~\cite{Bergmann2016}, which is 1.6 standard deviations higher than our result. This discrepancy suggests that the $P_{KK}$ values for the two neighboring nuclides still appear to deviate from the $1/Z^2$ scaling, differing from the consistency previously inferred by Bergmann \textit{et al}. Our experiment is a high-energy-resolution spectroscopy measurement, whereas Bergmann \textit{et al}. employed a  low-energy-resolution coincidence method. The high energy resolution enables direct resolution of the $K$ X-ray hypersatellites from the diagram lines. This makes the data analysis straightforward and also ensures that the observed X-rays originate from the decay of $^{55}\text{Fe}$, not from minor source contaminants or fluorescence from the source or adjacent materials. Furthermore, the single-detector method is independent of the fluorescence yields $\omega_{KL}$ and $\omega_{KM}$, which are required in coincidence detection schemes. Here $\omega_{KM}$ is the fluorescence yields for filling a single $K$-shell vacancy in the presence of a spectator vacancy in the $M$-shells. These fluorescence yields involve highly complex initial states~\cite{CHEN2015}, and discrepancies between the adopted and actual values of $\omega_{KL}$ and $\omega_{KM}$ could also introduce uncertainties. In addition, although the X-ray emission anisotropy of the second \textit{K} satellite emission following the detection of the hypersatellite by angular momentum correlation has been confirmed ~\cite{Bergmann2016}, the resulting uncertainty would also be excluded by the single-detector method. The present result may therefore be considered to have a smaller systematic error.

Theoretically, the pioneering calculation of $P_{KK}$ by Primakoff and Porter was performed within a non relativistic hydrogen like framework~\cite{Primakoff1953}. Subsequent refinements have introduced various relativistic corrections and different treatments of electron correlation effects~\cite{Law1973,Mukoyama1973,Suzuki1982,Intemann1985,Intemann1987}. And, these theoretical approaches yield widely scattered results. A summary of the methods for $^{55}$Fe is provided in Table\,I of Ref.~\cite{Campbell1991}, and Fig ~\ref{fig4} also presents these theoretical values. Among the different models, two leading theoretical treatments can be identified. The first is a full many-body approach based on a Dirac--Fock--Slater self-consistent-field (SCF) potential by Suzuki \textit{et al.} which predicts $P_{KK}=2.01\times 10^{-4}$~\cite{Suzuki1982}. The second is a semi-relativistic Coulomb-propagator (SRP) method by Intemann \textit{et al.} which gives {$P_{KK}=9.42\times 10^{-5}$~\cite{Intemann1985}. The former is about twice the latter. Although a fully relativistic SCF framework might be considered theoretically more complete, our experimental result falls between the SRP and SCF predictions, lying somewhat closer to the SRP.

The deviation of the $P_{KK}$ values for $^{55}\text{Fe}$, $^{54}\text{Mn}$,
and $^{65}\text{Zn}$ from the $1/Z^{2}$ scaling was attributed by
Campbell \textit{et al}. \cite{Campbell1991} to the fact that, in the
$^{54}\text{Mn}$ and $^{65}\text{Zn}$ cases, the EC decay does not end in
the ground state of the daughter nucleus but in an excited state, whereas
$^{55}\text{Fe}$ decays directly to the ground state of $^{55}\text{Mn}$.
Internal conversion (IC) decay of the excited nuclear state can eject the
remaining \textit{K}-shell electron, leaving a double \textit{K}-shell
vacancy. The \textit{K}-conversion coefficient $\alpha_K$ for both
$^{54}\text{Mn}$ and $^{65}\text{Zn}$ is on the order of $10^{-4}$ \cite{Lederer1978, WARD1986}, similar
to the $P_{KK}$ probability by shake-up and shake-off. However, Hindi \textit{et al.} \cite{Hindi1991} pointed out that IC can only contribute to $P_{KK}$ if it occurs before the vacancy created by
electron capture is filled. Given that the atomic \textit{K}-vacancy
lifetime ($\tau_a \sim 10^{-16}\,\text{s}$ \cite{Krause1979}) is approximately
$10^{3}$--$10^{4}$ times shorter than the daughter excitation lifetime
($\tau_n \sim 10^{-13}$--$10^{-12}\,\text{s}$ \cite{Lederer1978, WARD1986}), this mechanism rarely
produces an empty \textit{K} vacancy and merely creates additional single
\textit{K} vacancies. A second potential contribution is the Internal
Compton Effect (ICE). The nuclear excited state promptly decays by
$\gamma$-emission, and this $\gamma$ ray can cause Compton ionization of a
\textit{K}-shell electron. As the cross section steeply rises towards zero energy of the emitted electron, the scattered gamma loses only a tiny amount of energy of its typically 0.8--1~MeV energy in these cases. The probability for ICE in the Mn and Zn cases is also of the order of \(10^{-4}\), i.e.\ similar to the \(P_{KK}\) probability by shake-up and shake-off. However, ICE also contributes only to single \textit{K} vacancies due to the similar time-scale mismatch. It should be noted that, in a coincidence measurement lacking sufficient energy resolution, a normal \textit{K} x ray from the EC \textit{K} vacancy can coincide with another normal \textit{K} X-ray from a later IC or ICE \textit{K} vacancy, mimicking the hypersatellite–satellite coincidence signature from a double \textit{K} vacancy and leading to a spurious $P_{KK}$ count.

Given that the timescale arguments exclude contributions from IC and ICE, the observed deviation from the  $1/Z^{2}$ scaling likely originates from the $Q$-value dependence of $P_{KK}$ and the nuclear structure of the daughter state. 
Table~IV of Ref.~\cite{Hindi1992} summarizes 15 isotopes for which $P_{KK}$ has been measured, together with the theoretical predictions of the SRP and SCF models. Previous studies of EC decays with low $Q$ ($^{109}\mathrm{Cd}$~\cite{Eijk1979} $^{125}\mathrm{I}$~\cite{Hindi1992}, $^{139}\mathrm{Ce}$~\cite{Hindi199102}, $^{181}\mathrm{W}$~\cite{Eijk1982}) show that the measured $P_{KK}$ values are larger than both SRP and SCF theories predict. This is because both theories let the shake-off probability drop too quickly as the decay energy approaches threshold, without fully accounting for the phase-space effect that slows this decline. But for the three large-$Q$ ($Q/B_{K}>20$,  where $B_{K}$ is the \textit{K}-shell binding energy) EC-decay nuclides $^{54}\mathrm{Mn}$, $^{65}\mathrm{Zn}$, and $^{71}\mathrm{Ge}$, the measured $P_{KK}$ values also exceed or come close to the SCF predictions. This indicates that $P_{KK}$ may be governed not only by the $Q$ value but also by the nuclear structure of the daughter state.
Consequently, a more complete theoretical treatment of the shake-off amplitude across different $Q$ values and endpoint states of the daughter nucleus is needed to resolve the present discrepancies.
Experimentally, further $P_{KK}$ measurements with microcalorimeters for nuclides such as $^{54}\mathrm{Mn}$ and $^{71}\mathrm{Ge}$ should be prioritized, because these two EC nuclides share many common features with $^{55}\mathrm{Fe}$---similar $Z$, EC-dominant decay, large Q---yet their $P_{KK}$ values are significantly different.

\section{SUMMARY AND OUTLOOK}

This study presents the first experimental determination of the double \textit{K}-shell vacancy creation probability ($P_{KK}$) in the EC decay of $^{55}$Fe utilizing a microcalorimeter. By exploiting the detector's high energy resolution, we directly resolved the faint Mn \textit{K}-hypersatellite lines ($K\alpha_{2}^{h}$, $K\alpha_{1}^{h}$, $K\beta_{1,3}^{h}$) from the dominant diagram lines without resorting to complex spectral fitting or coincidence techniques. This work  therefore yields a smaller systematic error than previous studies. Based on the measured relative intensities and established fluorescence yields, we determined $P_{KK} = (1.22 \pm 0.09) \times 10^{-4}$. This value is consistent with most historical measurements and recent high-precision data, though a small discrepancy from the latest result reported by Bergmann \textit{et al}. remains.  While this measurement reinforces the established value for $^{55}$Fe, it does not resolve the persistent anomaly where neighboring nuclides like $^{54}$Mn and $^{65}$Zn exhibit significantly higher $P_{KK}$ values than predicted by the $1/Z^2$ scaling law, hinting at unaccounted nuclear structure effects or secondary processes, though the IC and ICE are unlikely due to timescale constraints. Theoretical analysis places our experimental finding between the predictions of SRP and SCF approach, suggesting that the SRP method offers a more accurate description for this specific decay regime.

Future investigations leveraging the capabilities of microcalorimeters should prioritize the re-examining of $P_{KK}$ for other EC isotopes, particularly $^{54}$Mn and $^{71}$Ge, which share similar atomic properties with $^{55}$Fe but display divergent probabilities. Applying this technique to these nuclei will be critical for distinguishing whether the deviations from theoretical scaling laws arise from specific daughter-nucleus excited states or from overlooked decay channels like IC and ICE. Additionally, future efforts should aim to integrate X-ray and Auger electron detection within a single microcalorimeter system~\cite{KOSSERT2022}. Such a unified scheme would allow for the direct measurement of double-vacancy fluorescence yields, thereby removing a major source of theoretical uncertainty and providing the thorough experimental benchmarks necessary to refine many-body quantum mechanical models of electron correlation and shake-off phenomena in nucleus decay.

\section {ACKNOWLEDGMENTS}
This work was supported by the National Key Research and Development Program of China under Grant No.2022YFA1602500; the National Natural Science Foundation of China under Grants under Grants No. 12375262, 12441507, 12275328, and U1732269. K. K., and R. S. acknowledge the support of the Chinese Academy of Sciences President's International Fellowship Initiative (PIFI) under Grants No. 2022VMC007, and 2026PD0037, respectively. We are grateful to the crew of the Low Energy high-intensity heavy ion Accelerator Facility (LEAF) for the experimental support.

\bibliography{Ref}      
\end{document}